\documentclass[RNAAS]{aastex62}

\begin{document}

\title{A Reassessment of Families of Solutions to the Puzzle of Boyajian's Star}

\correspondingauthor{Jason T.\ Wright}
\email{astrowright@gmail.com}
\author[0000-0001-6160-5888]{Jason T.\ Wright}
\affil{Department of Astronomy \& Astrophysics \\ and \\ Center for Exoplanets and Habitable Worlds \\ 525 Davey Laboratory, 
The Pennsylvania State University, 
University Park, PA, 16802, USA}



\section{Background and Recent Work}
\citet{Wright16} described several families of plausible solutions to the puzzle of Boyajian's Star \citep{WTF}, and ranked some of them by their subjective plausibility.

Since then, theoretical work has increased some families' plausibility, especially: \citet{Metzger17} and \citet{Wyatt2008} who describe models involving circumstellar material, and \citet{Metzger17} and \citet{Foukal17}, who describe models of intrinsic variations. 

Observationally, \citet{Meng17} and \citet{Davenport17} find that the long-term dimming is stronger in the blue and perhaps more consistent with circumstellar material than typical interstellar dust. \citet{Simon17} find that the long-term brightness variations are not monotonic (i.e.\ the star brightens and dims on timescales of years).

Most significantly, \citet{Boyajian18} and \citet{Deeg18} find that the dips are much stronger in the blue and consistent with attenuation by sub-micron dust grains, implying a very short orbital lifetime for the dust, if it is circumstellar. \citet{Boyajian18} also find that the stellar spectrum shows no increase in the column density of neutral (\ion{Na}{1} D) or ionized (\ion{Ca}{2} H \& K) gas during a dip. Finally, \citet{Boyajian18} confirm the lack of strong radial velocity variation found by \citet{WTF}.

This work alters my subjective plausibilities of several families of solutions. 

\section{Reassessed Families of Solutions}

{\it Instrumental effects} are now not worth any consideration: instruments other than {\it Kepler} have seen the dips. {\it Pulsations} are less likely now because of the constancy of in-dip RVs. Explanations involving only opaque objects such as {\it stars, planets, swarms of asteroids}, or {\it alien megastructures} \citep{TabbysStar} are now ruled out.

{\it Polar spots} continue to be unlikely. The {\it Solar System cloud} model still needs development \citep[but see][]{Katz17}.

All models involving {\it interstellar dust} are supported by the dips' colors, but now must grapple with the lack of increased gas column during dips. This is not necessarily a fatal problem: on the very small length scales required to produce the dips, dust and gas in the ISM may be uncoupled. Further, the ISM sodium absorption lines are saturated and so are relatively insensitive to additional neutral gas at the same RV as the known clouds. The calcium lines do not appear to be saturated, but might not be expected to trace the dust well, anyway.

The {\it post-merger return to normal} hypothesis is now plausible thanks to the work of \citet{Metzger17}, but still requires development to predict the color and spectral changes one expects to see, as well as an explanation for the secular {\it brightening}. 

{\it Intrinsic variability} now has some theoretical support \citep{Foukal17}, and appears to be consistent with the multiband photometry \citep{Sheikh16,Foukal17b,Boyajian18}, although a mechanism and explanation for its rarity remain elusive \citep[but see][]{Lacki17}. It is also consistent with the lack of significant additional gas absorption during a dip.

The measured reddening may favor {\it circumstellar material}, and thus moves this family of solutions up in plausibility, however these models still struggle with the lack of observed NIR excess and the short lifetime implied by the small dust grains found by \citet{Boyajian18} and \citet{Deeg18}. Additionally, the lack of gas accompanying the very fine dust would seem to argue against an ``exocometary'' origin for the dust. A quantitative calculation of the amount of gas and the dust replenishment rate in these scenarios would be helpful. It may be more likely that the material is produced collisionally from ``dry'' asteroids at large distances from the star.

Finally, the \cite{Wright16} hypothesis of an {\it intervening black hole disk} appears to still be in play, and should be investigated. In particular, such dust might be characteristic of circumstellar grains, and the lack of accompanying gas could be due to the unusual abundance patterns of such a disk, or it could be so cold that all of its gas has condensed onto grains. The dust in this hypothesis is far from any source of strong, radiation and so is consistent with the continued lack of IR excess, and the persistence of the dust despite its small typical grain size.

\acknowledgments

I thank Josh Peek, Eric Mamajek, and Tabetha Boyajian for useful discussions.

\bibliography{references}

\begin{thebibliography}{}
\expandafter\ifx\csname natexlab\endcsname\relax\def\natexlab#1{#1}\fi
\providecommand{\url}[1]{\href{#1}{#1}}

\bibitem[{{Boyajian} {et~al.}(2016){Boyajian}, {LaCourse}, {Rappaport},
  {Fabrycky}, {Fischer}, {Gandolfi}, {Kennedy}, {Korhonen}, {Liu}, {Moor},
  {Olah}, {Vida}, {Wyatt}, {Best}, {Brewer}, {Ciesla}, {Cs{\'a}k}, {Deeg},
  {Dupuy}, {Handler}, {Heng}, {Howell}, {Ishikawa}, {Kov{\'a}cs}, {Kozakis},
  {Kriskovics}, {Lehtinen}, {Lintott}, {Lynn}, {Nespral}, {Nikbakhsh},
  {Schawinski}, {Schmitt}, {Smith}, {Szabo}, {Szabo}, {Viuho}, {Wang},
  {Weiksnar}, {Bosch}, {Connors}, {Goodman}, {Green}, {Hoekstra}, {Jebson},
  {Jek}, {Omohundro}, {Schwengeler}, \& {Szewczyk}}]{WTF}
{Boyajian}, T.~S., {LaCourse}, D.~M., {Rappaport}, S.~A., {et~al.} 2016,
  \mnras, 457, 3988

\bibitem[{{Boyajian} {et~al.}(2018){Boyajian}, {Alonso}, {Ammerman},
  {Armstrong}, {Asensio Ramos}, {Barkaoui}, {Beatty}, {Benkhaldoun}, {Benni},
  {Bentley}, \& et~al.}]{Boyajian18}
{Boyajian}, T.~S., {Alonso}, R., {Ammerman}, A., {et~al.} 2018, ArXiv e-prints,
  arXiv:1801.00732

\bibitem[{{Davenport} {et~al.}(2017){Davenport}, {Covey}, {Clarke}, {Laycock},
  {Fleming}, {Boyajian}, {Montet}, {Shiao}, {Million}, {Wilson}, {Olmedo},
  {Mamajek}, {Olmedo}, {Chavez}, \& {Bertone}}]{Davenport17}
{Davenport}, J.~R.~A., {Covey}, K.~R., {Clarke}, R.~W., {et~al.} 2017, ArXiv
  e-prints, arXiv:1712.04948

\bibitem[{{Deeg} {et~al.}(2018){Deeg}, {Alonso}, {Nespral}, \&
  {Boyajian}}]{Deeg18}
{Deeg}, H.~J., {Alonso}, R., {Nespral}, D., \& {Boyajian}, T. 2018, ArXiv
  e-prints, arXiv:1801.00720

\bibitem[{{Foukal}(2017{\natexlab{a}})}]{Foukal17}
{Foukal}, P. 2017{\natexlab{a}}, \apjl, 842, L3

\bibitem[{{Foukal}(2017{\natexlab{b}})}]{Foukal17b}
---. 2017{\natexlab{b}}, Research Notes of the American Astronomical Society,
  1, 52

\bibitem[{{Katz}(2017)}]{Katz17}
{Katz}, J.~I. 2017, ArXiv e-prints, arXiv:1705.08377

\bibitem[{{Lacki}(2016)}]{Lacki17}
{Lacki}, B.~C. 2016, ArXiv e-prints, arXiv:1610.03219

\bibitem[{{Meng} {et~al.}(2017){Meng}, {Rieke}, {Dubois}, {Kennedy}, {Marengo},
  {Siegel}, {Su}, {Trueba}, {Wyatt}, {Boyajian}, {Lisse}, {Logie}, {Rau}, \&
  {Vanaverbeke}}]{Meng17}
{Meng}, H.~Y.~A., {Rieke}, G., {Dubois}, F., {et~al.} 2017, \apj, 847, 131

\bibitem[{{Metzger} {et~al.}(2017){Metzger}, {Shen}, \& {Stone}}]{Metzger17}
{Metzger}, B.~D., {Shen}, K.~J., \& {Stone}, N. 2017, \mnras, 468, 4399

\bibitem[{{Sheikh} {et~al.}(2016){Sheikh}, {Weaver}, \& {Dahmen}}]{Sheikh16}
{Sheikh}, M.~A., {Weaver}, R.~L., \& {Dahmen}, K.~A. 2016, Physical Review
  Letters, 117, 261101

\bibitem[{{Simon} {et~al.}(2017){Simon}, {Shappee}, {Pojmanski}, {Montet},
  {Kochanek}, {van Saders}, {Holoien}, \& {Henden}}]{Simon17}
{Simon}, J.~D., {Shappee}, B.~J., {Pojmanski}, G., {et~al.} 2017, ArXiv
  e-prints, arXiv:1708.07822

\bibitem[{{Wright} {et~al.}(2016){Wright}, {Cartier}, {Zhao}, {Jontof-Hutter},
  \& {Ford}}]{TabbysStar}
{Wright}, J.~T., {Cartier}, K.~M.~S., {Zhao}, M., {Jontof-Hutter}, D., \&
  {Ford}, E.~B. 2016, \apj, 816, 17

\bibitem[{{Wright} \& {Sigurdsson}(2016)}]{Wright16}
{Wright}, J.~T., \& {Sigurdsson}, S. 2016, \apjl, 829, L3

\bibitem[{{Wyatt}(2008)}]{Wyatt2008}
{Wyatt}, M.~C. 2008, \araa, 46, 339

\end{thebibliography}

\end{document}